\documentclass[
 aps,
 11pt,
 final,
 notitlepage,
 oneside,
 twocolumn,
 tightenlines,
 showkeys,
 nobibnotes,
 nofootinbib,
 superscriptaddress,
 noshowpacs,
 centertags,
 floatfix]
{revtex4}

\def\squareforqed{\hbox{\rlap{$\sqcap$}$\sqcup$}}

\def\sq{\ifmmode\squareforqed\else{\unskip\nobreak\hfil
\penalty50\hskip1em\null\nobreak\hfil\squareforqed
\parfillskip=0pt\finalhyphendemerits=0\endgraf}\fi}

\def\degr{\hbox{$^\circ$}}

\def\utw{\smash{\rlap{\lower5pt\hbox{$\sim$}}}}

\def\udtw{\smash{\rlap{\lower6pt\hbox{$\approx$}}}}

\def\farcs{\hbox{$\,.\!\!^{\prime\prime}$}}

\def\diameter{{\ifmmode\mathchoice
{\ooalign{\hfil\hbox{$\displaystyle/$}\hfil\crcr
{\hbox{$\displaystyle\mathchar"20D$}}}}
{\ooalign{\hfil\hbox{$\textstyle/$}\hfil\crcr
{\hbox{$\textstyle\mathchar"20D$}}}}
{\ooalign{\hfil\hbox{$\scriptstyle/$}\hfil\crcr
{\hbox{$\scriptstyle\mathchar"20D$}}}}
{\ooalign{\hfil\hbox{$\scriptscriptstyle/$}\hfil\crcr
{\hbox{$\scriptscriptstyle\mathchar"20D$}}}}
\else{\ooalign{\hfil/\hfil\crcr\mathhexbox20D}}%
\fi}}

\begin{document}
\selectlanguage{russian}



%


\title{SUPERFAST LINE PROFILE VARIATIONS IN THE SPECTRA OF OBA STARS III: A0 STAR $\alpha^2\,$CVn, NEW RESULTS}

\author{\firstname{A.~F.}~\surname{Kholtygin}}
 \email{afkholtygin@gmail.com}
 \affiliation{Saint Petersburg State University, Saint Petersburg, 199034 Russia}

\author{\firstname{A.~V.}~\surname{Moiseeva}}
 \affiliation{Special Astrophysical Observatory, Russian Academy of Sciences, Nizhnii Arkhyz, 369167 Russia}

\author{\firstname{I.~A.}~\surname{Yakunin}}
 \affiliation{Saint Petersburg State University, Saint Petersburg, 199034 Russia}
 \affiliation{Special Astrophysical Observatory, Russian Academy of Sciences, Nizhnii Arkhyz, 369167 Russia}

\author{\firstname{S.}~\surname{Hubrig}}
\affiliation{Leibniz-Institut fur Astrophysik Potsdam (AIP), Germany}

\begin{abstract}
This work is a continuation of the studies of the ultrafast variability of line profiles in the spectra of early-type stars. Line profile variations (LPVs) in the spectrum a chemically peculiar 
A0Vp star $\alpha^2\,$CVn are investigated using the January 6, 2020 observations carried out with the 6-meter BTA telescope at Special Astrophysical observatory (SAO) of 
the Russian Academy of Sciences (RAS) equipped with the MSS spectrograph. Regular short-term periodic variations of the H$_\beta$, Fe\,II, and Cr\,II lines were detected with periods ranging 
from $\sim\!$4 to $\sim\!$140 minutes. The magnetic field of the star was determined for all observations. The average measured longitudinal magnetic field component over the entire duration 
of observations is about $\approx$600\,G, which is close to the value expected from the well-known magnetic field phase curve. 
\end{abstract}


\maketitle

\section{INTRODUCTION}

Although line profile variations in the spectra of OBA stars are well-studied at time scales of hours to days \cite{Kaper-1997,Kholtygin-2003a,Dushin-2013}, 
the profile variations at minute and second scales remain practically unexplored until now.

The discovery of fast variations in the profiles of lines Si\,II and Fe\,II in the spectra of an A0 supergiant HD\,92207 at time scales of
1Ц2 minutes by Hubrig et al.~\cite{Hubrig-2014} prompted us to investigate the superfast spectral variations of early-type stars at minute time scales.

In order of investigate whether the short-term line profile variations in the spectra of OBA-stars are a common phenomenon, we analyzed such variations with a temporal resolution of 
minutes and fractions of minutes using the SCORPIO low-resolution spectrograph (focal reducer) of the 6-meter telescope of SAO RAS (Afanasiev and Moiseev 2005,~\cite{Af-2005}), as well 
as the FORS\,2 spectropolarimeter at the 8-meter VLT (Antu) telescope.

A review of observations carried out during the program of looking for the superfast LPVs in the spectra of OBA stars is presented by  
Batrakov et al.~\cite{Batrakov-2020} and Tsiopa et al.~\cite{Tsiopa-2020}.

Kholtygin et al.~\cite{Kholtygin-2017} presented the results of a study of superfast variations in the spectra of the HD\,93521 (O9.5III) star based on the 2015 BTA (SAO RAS) observations. 
Regular variations with periods of 4Ц5 and 32Ц36 minutes were detected. An analysis of the line profiles in the spectrum of Be-star $\lambda$\,Eri obtained with the FORS2 
spectropolarimeter by Hubrig et al.~\cite{Hubrig-2017} showed the presence of variations in the longitudinal magnetic field component with a period of 13.6 minutes. Such variations were
also detected in the Balmer and He\,I line profiles.

Batrakov et al. \cite{Batrakov-2019,Batrakov-2020} and Kholtygin et al.~\cite{Kholtygin-2018} presented the results of searching for fast variations in the spectra of the slowly rotating
$\rho\,$Leo (BIa) supergiant. Regular short period variations of the H and He lines were detected with periods ranging from 2 to 90 minutes. The same star was observed in OctoberЦNovember, 2019 
with the 1.25-meter telescope of the Crimean Astronomical Station of Sternberg Astronomical Institute  (SAI) of Moscow State University (MSU,~\cite{Kholtygin-2020a}). Short-term line profile 
variations were discovered at times scales of 15Ц25 minutes.

An analysis of the spectra of the A2 III giant $\gamma\,$UMi obtained in January 2017 with the BTA equipped with the SCORPIO spectrograph showed
the presence of harmonic components in the line profile variations with periods in the 10Ц65 minute interval~\cite{Tsiopa-2020}.

Of special interest is an investigation of a chemically peculiar and magnetic standard A0Vp star $\alpha^2\,$CVn. Despite the large number of 
publications\footnote{1637 ADS citations up to April 22, 2020} dedicated to the study of this star, its short-term variations have practically not been investigated.
One may mention only the study by Kuvshinov and Plachinda~\cite{Kuvshinov-1983}, where the authors investigate variations of the line profiles in the H and K Ca\,II line
center. They report a detection of irregular variations in the profiles of these lines on time scales ranging from minutes to hours.

In a recent paper by Kholtygin et al.~\cite{Kholtygin-2020} authors analyze $\alpha^2\,$CVn observations carried out on January 20/21, 2015 using the BTA telescope equipped with the SCORPIO 
spectrograph. Short-term regular variations in the Balmer lines and He lines with periods ranging from about $\sim\!$30 to 135 minutes were detected. Using window Fourier transform,
we detected quasi-regular transient profile variations in the Balmer lines with 3Ц6 minute periods.

In this paper we analyze the $\alpha^2\,$CVn BTA observations with a temporal resolution of 2Ц3 minutes carried out on January 6, 2020. The paper is organized as follows. Section~\ref{s.Obs} 
describes the observations and reduction of spectra. Section~\ref{s.LPV} is dedicated to the variation analysis of the line profiles. Measurements of the stellar magnetic field are presented in 
Section ~\ref{s.MagnField}, and a discussion of the obtained results is given in Section~\ref{s.Disc}. Our conclusions are outlined in Section~\ref{s.concl}.

\begin{figure*}[]
 \setcaptionmargin{5mm} \onelinecaptionstrue \captionstyle{normal}
 \includegraphics[scale=1.3]{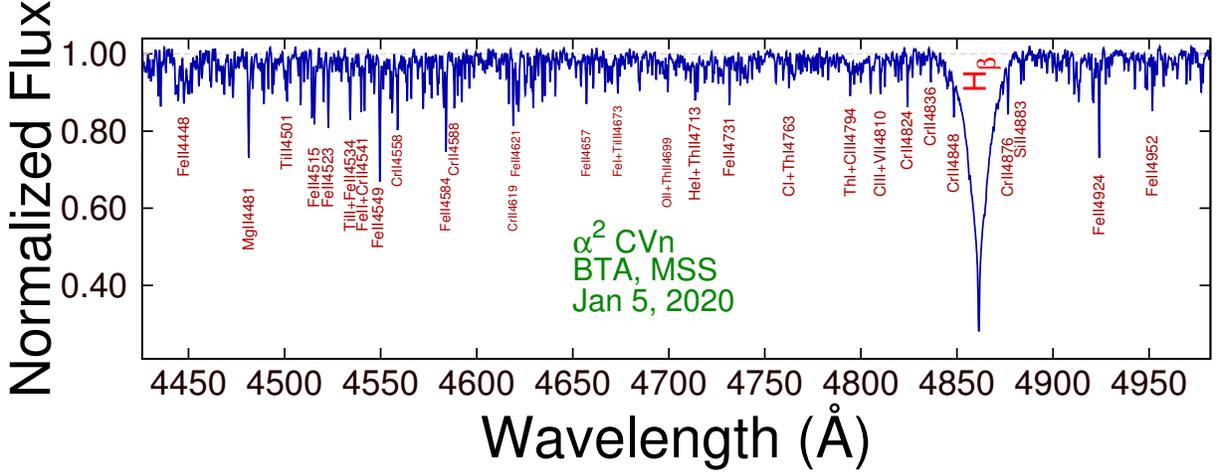}   
 \caption{Average normalized spectrum of $\alpha^2$\,CVn. Marked are the lines investigated for profile variations. 
          }
 \label{Fig.FullSp_alf2CVn}
\end{figure*}

\section{OBSERVATIONS AND REDUCTION OF SPECTRA}
\label{s.Obs}

The chemically peculiar star $\alpha^2\,$CVn (HD112413) is a magnetic standard. Its effective temperature $T_{\mathrm{eff}}=11320\pm 600$~K~\cite{Sikora-2019a}
is higher than that typical for A0 main sequence stars ($\sim\!$9600~K). A rather unusual peculiarity of the $\alpha^2\,$CVn star is the extreme weakness of its X-ray emission.
Observations of this star by the Chandra and XMM satellites show that its X-ray flux $\log L_{\mathrm{X}} <26.0$~erg/s, which is 3Ц4 orders of magnitude smaller than the typical 
values for $\alpha^2\,$CVn-type stars~\cite{Robrade-2011}.

\begin{figure}[ht!]
 \setcaptionmargin{5mm} \onelinecaptionstrue \captionstyle{normal}
 \includegraphics[scale=0.40]{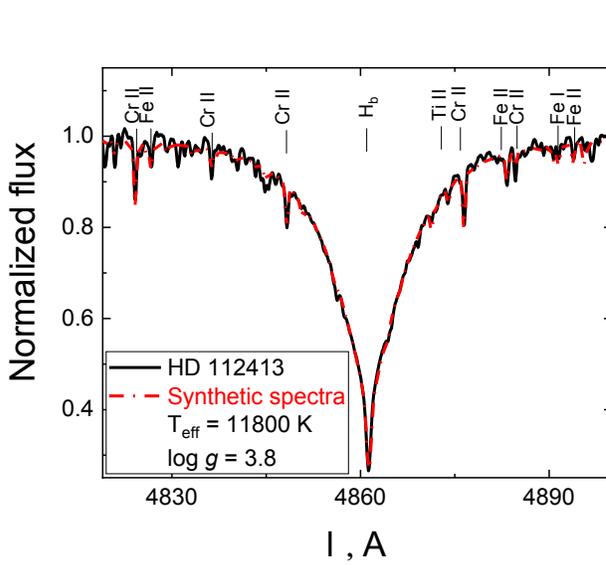}   
 \caption{\small Synthetic H$_{\beta}$ line profile 
         }                                   
 \label{Fig.SyntProf_Hbeta}                        
\end{figure}

The radial velocity of $\alpha^2\,$CVn varies significantly on time scales exceeding several months. The radial velocity $V_\mathrm{rad}$ variations possibly indicate a binarity
of $\alpha^2\,$CVn and a presence of a low-mass companion (Romanyuk and Semenko, 2007~\cite{Romanyuk-2007}). Radial velocity estimates based on SAO RAS observations show that
the variation period $V_\mathrm{rad}$ is probably $\approx\!$60 days or more.

\begin{figure}[ht!]
 \setcaptionmargin{5mm} \onelinecaptionstrue \captionstyle{normal}
 \includegraphics[angle=270,scale=0.30]{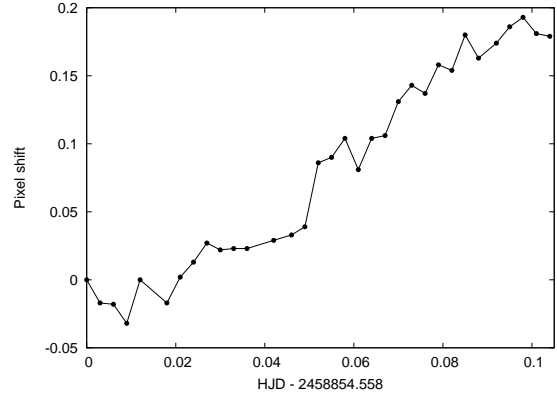}   
 \caption{\small Spectral shifts in the pixel scale of the CCD detector of the MSS spectrograph.
          }
 \label{Fig.SpectraShifts}
\end{figure}

Observations of the star were carried out with the 6-meter BTA telescope within the framework of the УMicrovariability in OB starsФ program (principal investigator is A.F. Kholtygin, SPbSU) 
using the MSS spectrograph (Panchuk et al. 2014~\cite{Panchuk-2014}) equipped with a circular polarization analyzer (Chountonov et al.~\cite{Chountonov-2007,Chountonov-2016}) and an 
image slicer. In order to eliminate the instrumental polarization, circularly polarized spectra were taken in pairs in two phase plate positions changing the polarization plane by 90\degr.
This procedure allows us to consequently obtain an orthogonally polarized signal at the same pixels of the CCD-detector. 

In order to control the obtained values, the spectra of 
magnetic field standards (usually these are stars with a well known magnetic phase curve) and zero polarization standards were obtained.

\begin{figure}[ht!]
 \setcaptionmargin{5mm} \onelinecaptionstrue \captionstyle{normal}
 \includegraphics[scale=0.55]{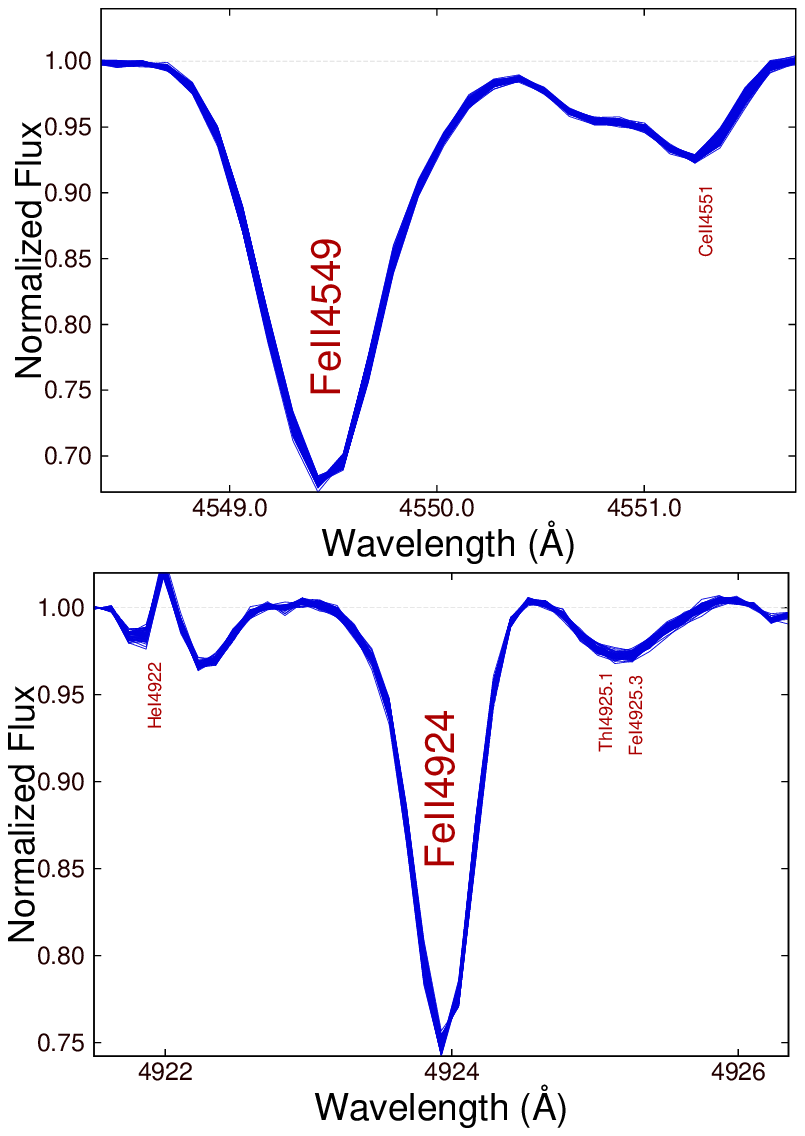}   
 \caption{\small FeII\,4549 (a) and FeII\,4924 (b) line profiles for all 71 spectra of $\alpha^2\,$CVn. The variations of individual profiles
                  are small enough to blend into one profile, overlapping each other.
          }                                   
 \label{Fig.LPValf2CVn_FeII}                        
\end{figure}

\begin{figure}[ht!]
 \setcaptionmargin{5mm} \onelinecaptionstrue \captionstyle{normal}
 \includegraphics[scale=0.60]{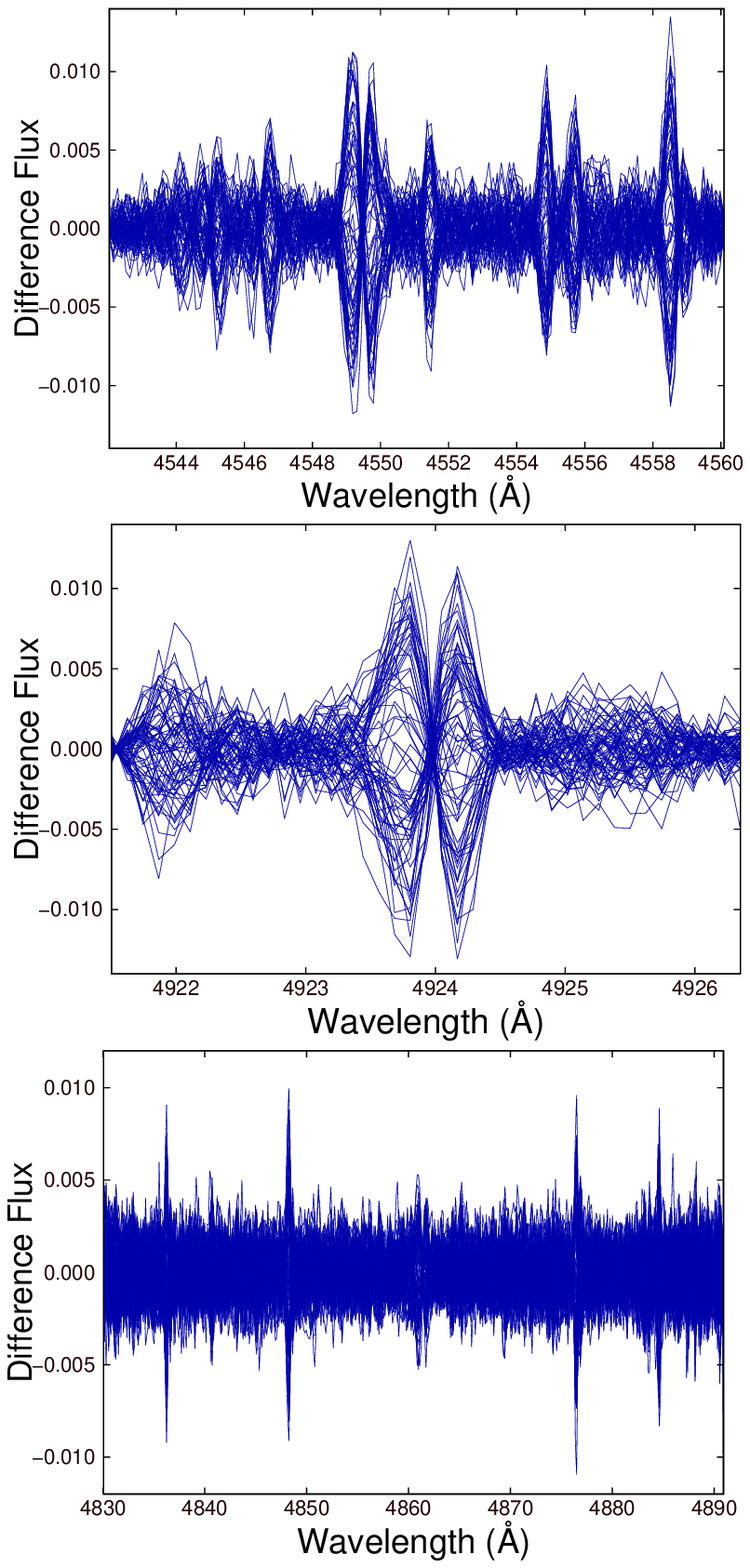}   
 \caption{\small Deviations of the FeII\,4549, FeII\,4924, and H$_{\beta}$ line profiles from the corresponding average profiles (top to bottom).
          }
 \label{Fig.VarySpFeIIHbeta}
\end{figure}

The primary reduction and spectra extraction were performed with ESO MIDAS system using the ZEEMAN context (Kudryavtsev et al.~\cite{Kudryavtsev-2000}). The following standard procedures 
were used for spectra reduction: CCD array bias subtraction, scattered light subtraction, wavelength calibration, after which the one dimensional spectra were normalized to the solar system
centroid. When performing wavelength calibration for the series of spectra, we used the nearest in time ThAr lamp spectrum. To normalize the spectra to
the continuum, the latter was interpolated by a cubic spline using the "spectool" package in the IRAF\footnote{\tiny http://iraf.noao.edu/projects/spectroscopy/spectool/spectool.html} environment.
The process of MSS observation reduction is described in more detail by Kudryavtsev et al.~\cite{Kudryavtsev-2000} and Semenko et al.~\cite{Semenko-2017}.

The spectral resolution amounts to about 15000 (slit size 0\farcs5), the range of registered wavelengths is 4425Ц4982\,\AA, the average S/N is approximately 700.
Over the course of our observations with a total time of 156 minutes, we obtained 71 spectra of the star with simultaneously recorded left and right circular
polarizations, with an exposure of 90 seconds each. The temporal resolution of one spectrum with account for the CCD detector readout is about 130 seconds.

Figure~\ref{Fig.FullSp_alf2CVn} shows the total spectrum of the star $\alpha^2\,$CVn normalized to the continuum and averaged over all 71 MSS observations:
\begin{equation}
\label{Eq.FullProf}
F(\lambda)= F_{\mathrm{L}}(\lambda) + F_{\mathrm{R}}(\lambda),
\end{equation} 
where $F_{\mathrm{L}}(\lambda)$ is the average intensity of the counter clockwise polarized spectral component normalized to the continuum, and $F_{\mathrm{R}}(\lambda)$ is that of the clockwise
polarized component.

\subsection{Model H$_{\beta}$ Line Profile}
\label{ss.SyntLineProfHbeta}

To refine the parameters of $\alpha^2\,$CVn  and determine the radial velocity of the star we model its synthetic spectrum using the ATLAS9 LTE models by Kurucz~\cite{Kurucz-1993}. 
The derived synthetic H$_{\beta}$ line profile in comparison with the observed profile averaged over all the recorded spectra is presented in Fig.~\ref{Fig.SyntProf_Hbeta}. Model parameters were 
taken from Kochukhov et al.~\cite{Kochukhov-2002}. For better agreement between the model and observed line profile, the gravity acceleration logarithm $\log g=4.02$ given by 
Kochukhov et al.~\cite{Kochukhov-2002} in their Table 3 was reduced to 3.8. Note that the model spectra describes both the H$_{\beta}$ and faint CrII\,4848.24 and CrII\,\,4876.39731 
line profiles with a very high accuracy.

The radial velocities $V_\mathrm{rad}$ determined by comparing model and observed H$_{\beta}$ line profiles from the obtained $\alpha^2\,$CVn spectra are presented in the last column of
Table~\ref{Table.FourSp}.

\subsection{Instrumental Effects}
\label{ss.InstrEff}

When analyzing fast spectral variations one needs to carefully take into account the instrumental errors, which are unavoidable due to the thermal and vibration instability of the spectrograph. 
The problem of the MSS positional stability has been multiply investigate (see, for example, papers~\cite{ChountonovNajdenov2009,kloch2008,Panchuk-2014}).  
The pattern of the long-term spectral shift over an interval of about 1.5 days is shown by Panchuk et al.~\cite{Panchuk-2014}. In addition to the long-term УdeparturesФ of the spectrum, 
rapid variations of its position were also noted. The effects related to the optomechanical instability of the MSS as well as methods of their minimization are described by 
Klochkova~\cite{kloch2008}. The results of a study of thermal stability are presented in a SAO preprint by Panchuk et al.~\citep{prep212}.

Shifts of 10 microns/hour were detected for the spectral line positions. The typical changes of the position of the spectrum with time in the current MSS configuration includes those a 
circular polarization analyzer and an image slicer as it presented in Fig. 1 in a paper by Chountonov and Najdenov~\cite{ChountonovNajdenov2009}. As is evident from the figure  
the drift is linear and does not exceed 0.3 pixels in 2.5 hours. Considering everything mentioned above, the average accuracy of determining the spectral line radial velocities 
from the MSS spectra is equal to 1.5Ц2 km\,s$^{-1}$.                         

To determine the value of the instrumental drift of the spectral lines during the current observation series we performed a cross-correlation analysis of each consecutive spectrum 
comparing it with the first one. The results of spectral shift measurements in the pixel scale of the CCD detector are presented in Fig.~\ref{Fig.SpectraShifts}. As is evident from the figure, 
both a long-term linear drift and short-term variations are present. The total shift between the first and last spectra of the series amounts to 0.18 pixels. Additionally, we used the 
calibration lamp spectra obtained in the beginning and end of the series, with a difference of 2 hours 40 minutes. Cross-correlation showed a relative shift between two spectra of 0.194 pixels. 
The detected long-term radial velocity variations are mainly instrumental and were taken into account in further analysis of the $\alpha^2\,$CVn spectra.

\section{LINE PROFILE VARIATIONS}
\label{s.LPV}

When analyzing line profile variations in the spectrum of a star, it is crucial to consider lines of sufficient depth and without significant blending. Based on these principles, 
we selected 6 lines: MgII\,4481, FeII\,4549, FeII\,4584, CrII\,4824, FeII\,4924, and  H$_\beta$. 

For illustrative purposes, Fig.~\ref{Fig.LPValf2CVn_FeII} shows all the obtained FeII\,4549 and FeII\,4924 line profiles in the spectrum of $\alpha^2\,$CVn. Their deviations from the average
profiles are shown in Fig.~\ref{Fig.VarySpFeIIHbeta}. As is evident from the figure, the amplitude of the variations does not exceed 1.5\%. The changes in the line profiles may be related
to the velocity field variations in the atmosphere of the star, for instance, due to non-radial pulsations.

It is noticeable that the amplitude of the H$_\beta$ line profile variations (Fig.~\ref{Fig.VarySpFeIIHbeta}, bottom) is significantly smaller than the amplitudes of the FeII\,4549 and
FeII\,4924 line profile variations, as well as those of the narrow lines Fe\,II and Cr\,II inside the H$_\beta$ line profile (see Figs.~\ref{Fig.FullSp_alf2CVn}-\ref{Fig.SyntProf_Hbeta}). 
We can assume that such differences are related to the inhomogeneity of the radial distribution of elements (e.g., Silvester et al.~\cite{Silvester-2014}).

\begin{figure}[ht!]
 \setcaptionmargin{5mm} \onelinecaptionstrue \captionstyle{normal}
 \includegraphics[scale=0.65]{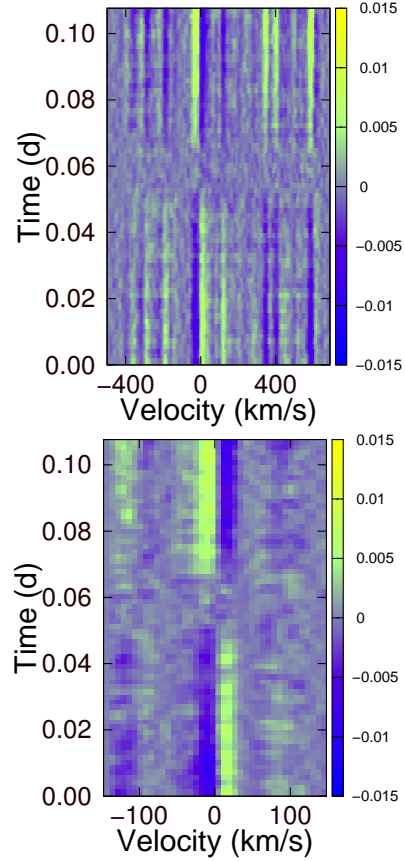}   
 \caption{\small Dynamic variation spectrum of the FeII\,4549 (top) and FeII\,4924 (bottom) line profiles. 
          }
 \label{Fig.dynSpFeII}
\end{figure}

\subsection{Regular Components of the Line Profile Variations}
\label{ss.regLPV}

In order to search for the regular components of the profile variations, let us determine the differential line profiles. Let $N$ spectra of the investigated object be
obtained as a result of the observations. Let $F_i(\lambda),~i=1, \dots, N$ denotes the continuum normalized flux in the $i$-th spectrum of the star at the wavelength $\lambda$.
Let $\overline{F}_i(\lambda)$ be the flux at wavelength $\lambda$ averaged over all observations.
The residual line profile is
\begin{equation}
\label{Eq.DiffProf}
d(\lambda)=F_i(\lambda) - \overline{F}_i(\lambda)
\end{equation} 
When analyzing subtractive profiles, instead of wavelengths, it is more convenient to use Doppler shifts $V$ from the laboratory wavelength $\lambda_0$ of the line 
$V=c(\lambda/\lambda_0-1)$,where $c$ is the speed of light.

If the spectra quality differs significantly, when computing the average and residual line profiles, one should use different profile weights $g_i$,
proportional to the squared signal-to-noise ratio in the continuum region near the line. However, the signal-to-noise ratios are close in all of our analyzed profiles and we
can therefore assume $g_i=1$.

Figure~\ref{Fig.dynSpFeII} shows the dynamic variation spectrum for the line profiles of FeII\,4549 and FeII\,4924 lines in the spectrum of $\alpha^2\,$CVn. Regular changes of the line 
profiles with time are evident. One can notice that the profile variations are coherent for both FeII\,4549 and the neighboring Ti\,II, Cr\,II, Fe\,I and Fe\,II lines, which indicates a common mechanism of 
their variations.

\begin{figure}[ht!]
 \setcaptionmargin{5mm} \onelinecaptionstrue \captionstyle{normal}
 \includegraphics[scale=0.85]{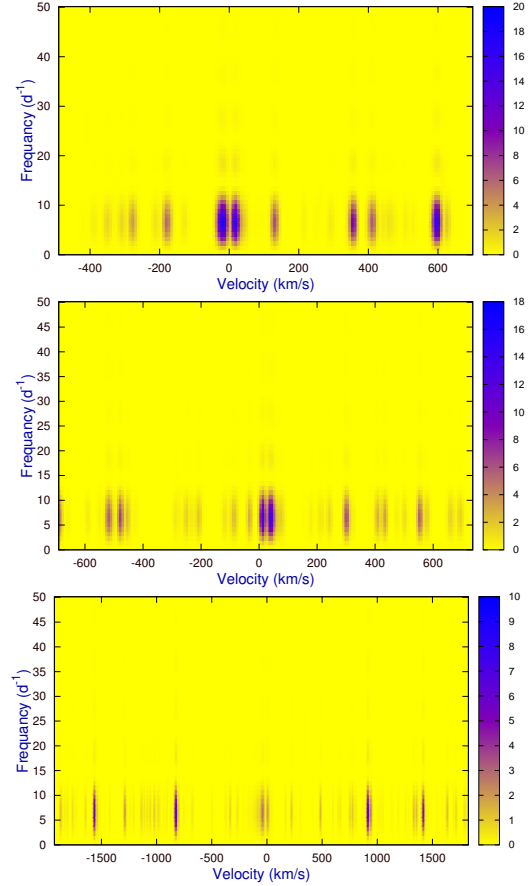}   
 \caption{\small Fourier spectra of the FeII\,4549, CrII\,4824 and H$_{\beta}$ line profile variations (top to bottom).
          }
 \label{Fig.FourSpLPValf2CVn}
\end{figure}

We carried out a search for periodic variation components in the line profiles of the $\alpha^2\,$CVn  spectrum using the CLEAN method by Roberts et al.~\cite{Roberts-1987}. 
Fourier spectra of the variations of the difference Fe\,II4549, CrII\,4824 and H$_{\beta}$ line profiles (periodograms) are presented in Fig.~\ref{Fig.FourSpLPValf2CVn}  for False Alarm 
Probability (FAP) value ~$\alpha=10^{-3}$.

Regular components exceeding the white noise periodogram counts corresponding to the selected FAP value are presented in the Fourier spectrum. Table~\ref{Table.FourSp} gives the 
detected frequencies and periods of the possible harmonic components of the variations for the analyzed line profiles for $\alpha=10^{-2} \mbox{---} 10^{-5}$.

  \begin{table*}[ht]
\centering
        \caption{\small Frequencies (in d$^{-1}$ column 2) and periods (in minutes, column 3) of the regular profile variation components in the spectrum of~$\alpha^2\,$CVn. 
                        The $+$ sign in columns 4Ц9 indicates that the component was detected in the line profile variations, and the $-$ sign shows that it was not. 
                        Column 10 gives the periods of profile variations reported in Kholtygin et al. (2020)~\cite{Kholtygin-2020}. The last column shows FAP values for the 
                        detected Fourier components. Components 1Ц4 have periods longer than the total time of observations.
                }
        \label{Table.FourSp}
{\small\begin{tabular}{ccccccccccc}     
\hline
є ~~&$\nu$~(d$^{-1}$)&$P\,$(мин)&Mg\,II4481&Fe\,II4549&FeII\,4584&CrII\,4824&H$_\beta$&FeII\,4924&\cite{Kholtygin-2020}&$\alpha$  \\ \hline 
  1~~ &$  5.57\pm 9.23$& $259\pm 429$    &  $+$&  $+$ &   $+$    &   $+$   & $+$      & $+$      &  $-$                &$10^{-5}$  \\
  2~~ &$  6.50\pm 9.23$& $222\pm 315$    &  $+$&  $+$ &   $+$    &   $+$   & $+$      & $+$      &  $-$                &$10^{-5}$  \\
  3~~ &$  7.42\pm 9.23$& $194\pm 241$    &  $+$&  $+$ &   $+$    &   $+$   & $+$      & $+$      &  $-$                &$10^{-5}$  \\ 
  4~~ &$  8.35\pm 9.23$& $172\pm 190$    &  $+$&  $-$ &   $-$    &   $+$   & $+$      & $-$      &  $-$                &$10^{-5}$  \\ 
\hline                                                                                                                    
  5~~ &$ 10.67\pm 9.23$& $135\pm 116$    &  $-$&  $+$ &   $-$    &   $-$   & $+$      & $-$      & $135\pm 164$        &$10^{-2}$  \\ 
  6~~ &$ 15.31\pm 9.23$& $ 94.0\pm 56.7$ &  $+$&  $+$ &   $-$    &   $-$   & $-$      & $-$      &  $-$                &$10^{-2}$  \\
  7~~ &$ 26.91\pm 9.23$& $ 53.5\pm 18.3$ &  $+$&  $+$ &   $-$    &   $-$   & $+$      & $-$      & $ 52 \pm  25$       &$10^{-2}$  \\ 
  8~~ &$ 86.31\pm 9.23$& $ 16.7\pm  1.8$ &  $-$&  $-$ &   $-$    &   $-$   & $+$      & $+$      & $17.7\pm 2.8$       &$10^{-2}$  \\
  9~~ &$111.37\pm 9.23$& $ 12.9\pm  1.1$ &  $-$&  $-$ &   $+$    &   $+$   & $+$      & $-$      &  $-$                &$10^{-2}$  \\
 10~~ &$288.64\pm 9.23$& $  4.99\pm 0.16$&  $+$&  $-$ &   $+$    &   $-$   & $-$      & $-$      &  4.5-5.6            &$10^{-2}$  \\ 
\hline                                                                                                                    
\end{tabular}                  
}
  \end{table*}

For an upper estimation of the Fourier spectrum regular component frequency error we used the expression $\Delta\nu < 1/T$~\cite{Vityazev-2001}
where $T=156\,$minutes is the total duration of observations. Fourier components with a frequency difference less than $\Delta\nu$
were considered as one component with a frequency corresponding to the average frequency of these components.

The $+$ sign in Table~\ref{Table.FourSp} indicates that the corresponding component is present in the Fourier spectrum, and the $-$ sign shows that is has not been
detected at the given FAP, although it may be registered at a higher one.

In order to exclude the components corresponding to random outliers of the periodogram, Table~\ref{Table.FourSp} includes only the Fourier spectrum components
detected in the profile variations of at least two lines. 

In the next to last column of the table we present the frequencies of the regular components of the profile variations in the spectrum of $\alpha^2\,$CVn obtained from SAO RAS BTA observations 
in 2015 with the SCORPIO spectrograph (Kholtygin et al.~\cite{Kholtygin-2020}). The corresponding FAP values are given for all of the found components in the last column of the table.

Components $\nu_1-\nu_4$ correspond to the periods $P$ which exceed the total duration of observations, $T=156\,$minutes. Further observations are needed to verify their reality.
Regular line profile variations in the spectra of OB stars with periods close to the periods of components $\nu_1-\nu_4$ with $P=3-6^h$ are most likely related to non-radial pulsations of 
the star in quadrupole ($l=2$) or higher pulsation modes accordingly by~Pamyatnykh~\cite{Pamyatnykh1999}. The detection of line profile variations in the short period region $P=15-140^m$ 
may indicate a presence of high mode of non-radial pulsations with $l=6-12$.

Three regular profile variation components $\nu_5$, $\nu_7$, and $\nu_8$ in the $\nu >10\,\mathrm{d}^{-1}$ frequency region correspond to (with account for errors) the components 
$135\pm 164\,$minutes, $52\pm 25\,$minutes, and $17.7\pm 2.8\,$minutes reported by Kholtygin et al.~\cite{Kholtygin-2020} which confirms that they are real.

Using the windowed Fourier transform by Kholtygin et al.~\cite{Kholtygin-2020} showed the presence in the LPVs in the spectrum of $\alpha^2\,$CVn of short-period regular components with
variable frequency in the 4.5Ц5.6 minute period interval. The detected by us component $\nu_{10}$ corresponds to this frequency interval.

Earlier we interpreted the presence of line profile variation components with such short periods in the spectrum of HD\,93521 as evidence for the presence of high modes of non-radial pulsations 
with $l=20Ц60$ (see Kholtygin et al.~\cite{Kholtygin-2017}). The same interpretation is possible for short-term LPVs in the spectrum of $\alpha^2\,$CVn.

\section{MAGNETIC FIELD OF $\alpha^2\,$CVn}
\label{s.MagnField}

All $\alpha^2\,$CVn spectra were obtained using a circular polarization analyzer, which allows us to estimate the magnetic field of the star. To increase the accuracy of the 
determined Stokes parameter V we used consecutive observations with different angles of the quarter wave plate:
\begin{equation}
\label{Eq.VdivI}
\frac{V}{I}= \frac{1}{2}\left\{\left(\frac{I^o-I^e}{I^o-I^e}\right)_{-45^{\circ}} -\left(\frac{I^o-I^e}{I^o-I^e}\right)_{+45^{\circ}}\right\}, 
\end{equation} 
where $I^o$ and $I^e$ are the ordinary and extraordinary beams, correspondingly.

To determine the magnitude of the longitudinal component of the magnetic field averaged over the whole stellar disk (the effective magnetic field, $B_\mathrm{e}$), we used two
methods: (1) a modified Babcock method based on measuring shifts of the centers of gravity (cog) of the circularly polarized line components (e.g., Borra and Landstreet~\cite{Borra-1973}), 
and (2) a regression method based on studying the circular polarization of spectral lines (e.g.~Hubrig et al.~\cite{Hubrig-2014}) using the standard relation:
\begin{equation}
\label{Eq.MFreg}
\frac{V}{I}= -\frac{g_{\mathrm{eff}} e }{4\pi m_\mathrm{e} c^2}\lambda^2\frac{1}{I}\frac{dI}{d\lambda} B_\mathrm{e}.
\end{equation} 
where $g_{\mathrm{eff}}$ is the effective Lande factor of a line, $\lambda$ is its wavelength, $e$ is the electron charge, and $m_\mathrm{e}$ is the electron mass. 
$I$ is the unpolarized line intensity, $dI/d\lambda$ is the wavelength derivative of Stokes $I$ parameter.

The values of $B_\mathrm{e}$ obtained by the cog ($B_e^{\mathrm{cog}}$) and regression ($B_e^{\mathrm{regr}}$) methods and the corresponding standard deviations are 
presented in Table~\ref{Table.MF_Vrad}.

  \begin{table*}[ht]
\centering
        \caption{\small Results of magnetic field and radial velocity measurements for~$\alpha^2\,$CVn.}
        \label{Table.MF_Vrad}
{\small 
\begin{tabular}{ccccccc}     \hline
          &Phase      &Phase      &Phase     &$B_e^{\mathrm{cog}},$&$B_e^{\mathrm{regr}},$&$V_{\mathrm{rad}},$\\  
MJD       &$P\!\!=\!5.49800$&$P\!\!=\!5.46939$&$P\!\!=\!5.43730$&G&  G                  &    $km\,s^{-1}$   \\  
\hline                                   
58854.058 & 0.7308    & 0.8219    & 0.8850   & $-662\pm 29$       &  $-628\pm 11$       &~~$-0.31\pm 2.00$ \\
58854.061 & 0.7314    & 0.8225    & 0.8855   & $-752\pm 30$       &  $-682\pm 10$       &~~$-0.24\pm 1.90$ \\
58854.064 & 0.7319    & 0.8230    & 0.8861   & $-677\pm 32$       &  $-621\pm 11$       &~~$-0.39\pm 1.90$ \\
58854.067 & 0.7324    & 0.8236    & 0.8866   & $-624\pm 31$       &  $-558\pm 11$       &~~$-0.36\pm 2.10$ \\
58854.070 & 0.7330    & 0.8241    & 0.8872   & $-732\pm 33$       &  $-659\pm 10$       &~~$-0.12\pm 1.80$ \\
58854.076 & 0.7341    & 0.8252    & 0.8883   & $-676\pm 40$       &  $-647\pm 11$       &~~$-0.30\pm 1.90$ \\
58854.079 & 0.7346    & 0.8258    & 0.8888   & $-684\pm 38$       &  $-599\pm 13$       &~~$-0.09\pm 2.10$ \\
58854.082 & 0.7352    & 0.8263    & 0.8894   & $-643\pm 31$       &  $-595\pm 11$       &~~$-0.09\pm 2.00$ \\
58854.085 & 0.7357    & 0.8269    & 0.8899   & $-616\pm 30$       &  $-574\pm 10$       &~~$-0.01\pm 2.00$ \\
58854.088 & 0.7363    & 0.8274    & 0.8905   & $-679\pm 28$       &  $-625\pm 10$       &~~$-0.04\pm 1.90$ \\
58854.091 & 0.7368    & 0.8280    & 0.8910   & $-679\pm 26$       &  $-635\pm 10$       &~~$ 0.06\pm 1.80$ \\
58854.094 & 0.7374    & 0.8285    & 0.8916   & $-766\pm 35$       &  $-667\pm 10$       &~~$ 0.08\pm 1.90$ \\
58854.100 & 0.7385    & 0.8296    & 0.8927   & $-693\pm 28$       &  $-660\pm 10$       &~~$-0.02\pm 2.10$ \\
58854.104 & 0.7392    & 0.8303    & 0.8932   & $-680\pm 29$       &  $-618\pm 10$       &~~$ 0.09\pm 2.00$ \\
58854.107 & 0.7397    & 0.8309    & 0.8938   & $-732\pm 33$       &  $-627\pm 10$       &~~$ 0.21\pm 2.00$ \\
58854.110 & 0.7403    & 0.8314    & 0.8943   & $-659\pm 31$       &  $-586\pm 11$       &~~$ 0.28\pm 1.90$ \\
58854.113 & 0.7408    & 0.8320    & 0.8949   & $-638\pm 32$       &  $-579\pm 11$       &~~$ 0.33\pm 1.80$ \\
58854.116 & 0.7414    & 0.8325    & 0.8954   & $-605\pm 30$       &  $-540\pm 10$       &~~$ 0.46\pm 1.90$ \\
58854.119 & 0.7419    & 0.8331    & 0.8960   & $-647\pm 33$       &  $-574\pm 10$       &~~$ 0.30\pm 2.10$ \\
58854.122 & 0.7425    & 0.8336    & 0.8965   & $-711\pm 41$       &  $-654\pm 11$       &~~$ 0.47\pm 2.00$ \\
58854.125 & 0.7430    & 0.8342    & 0.8971   & $-785\pm 33$       &  $-699\pm 12$       &~~$ 0.57\pm 2.00$ \\
58854.128 & 0.7435    & 0.8347    & 0.8976   & $-728\pm 36$       &  $-637\pm 12$       &~~$ 0.74\pm 1.90$ \\
58854.131 & 0.7441    & 0.8353    & 0.8982   & $-672\pm 36$       &  $-568\pm 11$       &~~$ 0.84\pm 1.80$ \\
58854.134 & 0.7446    & 0.8358    & 0.8988   & $-603\pm 34$       &  $-585\pm 11$       &~~$ 0.69\pm 1.90$ \\
58854.137 & 0.7452    & 0.8364    & 0.8993   & $-705\pm 34$       &  $-622\pm 11$       &~~$ 0.87\pm 2.10$ \\
58854.140 & 0.7457    & 0.8369    & 0.8999   & $-701\pm 33$       &  $-635\pm 12$       &~~$ 0.83\pm 2.00$ \\
58854.143 & 0.7463    & 0.8375    & 0.9004   & $-674\pm 32$       &  $-594\pm 12$       &~~$ 1.01\pm 2.00$ \\
58854.146 & 0.7468    & 0.8380    & 0.9010   & $-757\pm 30$       &  $-653\pm 11$       &~~$ 0.76\pm 1.90$ \\
58854.150 & 0.7475    & 0.8388    & 0.9019   & $-731\pm 34$       &  $-655\pm 11$       &~~$ 0.91\pm 1.80$ \\
58854.153 & 0.7481    & 0.8393    & 0.9024   & $-702\pm 31$       &  $-632\pm 12$       &~~$ 0.89\pm 1.90$ \\
58854.156 & 0.7486    & 0.8399    & 0.9030   & $-740\pm 28$       &  $-698\pm 11$       &~~$ 0.94\pm 2.10$ \\
58854.159 & 0.7492    & 0.8404    & 0.9035   & $-728\pm 30$       &  $-679\pm 11$       &~~$ 0.91\pm 2.00$ \\
58854.162 & 0.7497    & 0.8409    & 0.9041   & $-769\pm 33$       &  $-700\pm 13$       &~~$ 0.82\pm 2.00$ \\ \hline
\end{tabular}                           
}
\end{table*}

\section{DISCUSSION OF RESULTS}
\label{s.Disc}

\subsection{Influence of Instrumental Effects and Atmospheric Variations on the Line Profile Variations}
\label{sss.Infl_InstrAtm}

A rather important question is to what extent the regular profile variation components found in the spectrum of $\alpha^2\,$CVn  may be related to instrumental
effects and, in particular, to the oscillations of the telescope itself and the detector. An analysis of a large sample of the 6-m SAO RAS telescope observations
has shown that the oscillations of the telescope itself are most likely irregular and occur on time scales uncorrelated with the detected periods (the detailed
description of the BTA positional instability are given, for example, Klochkova et al. ~\cite{kloch2008}. With regard to the instrumental drift of the spectral lines discussed in 
section 2.2, the radial velocity variation analysis presented in Fig. 3 has shown that after the subtraction of the trend in $V_{\mathrm{rad}}$,  marginal (FAP level $\alpha >0.01$) 
period of about 100 minutes is present, which does not correspond to any of the periods given in Table~\ref{Table.FourSp}.

Note also that the coincidence of the detected line profile variation periods in the spectrum of $\alpha^2\,$CVn with those derived by Kholtygin et al.~\cite{Kholtygin-2020} using
the SCORPIO spectrograph with another set of intrinsic oscillations also indicates that these periods are real. The set of frequencies and profile variation periods obtained in the 
analysis of the line profiles of various OBA stars (see references in the introduction) differs significantly, which shows an absence of a connection between the profile variation periods derived
in this work and the specific features of the instruments used.

We should also note that some contribution to the line profile variations in the spectrum of the star may be attributed to sporadic variations of the size of the
turbulent disk of the star due to atmospheric fluctuations. It is difficult to expect however, that the line profile variations caused by the influence of atmospheric
fluctuations would be in any way regular. At the same time, the influence of such fluctuations on the irregular profile variations cannot be ruled out entirely, which should be 
taken into account during analysis. Additionally, this effect is minimized when using an image slicer, as we did in this case.

\subsection{Rotation Period of $\alpha^2\,$CVn and the Magnetic Field Phase Curve}
\label{ss.RotPeriod}

The rotation period of $\alpha^2\,$CVn has been determined as far back as by Farnsworth~\cite{Farnsworth1932}, $P=5.46939\,$d. Consequent studies have slightly refined the period. According 
to Sikora et al.~\cite{Sikora-2019b} $P=5.46913\,$d. Recently 27-day photometric observations of $\alpha^2\,$CVn by the TESS satellite have become available~\cite{TESS-2020}.

An analysis of these data by the CLEAN method gives $P\!=\!5.43730\!\pm\!0.470\,$d, which corresponds to the values obtained by other authors. To refine the period one must analyze longer 
series of observations. Due to some uncertainty in the determination of the rotation period of $\alpha^2\,$CVn we varied the period within the limits of its error.

\begin{figure}[ht!]
 \setcaptionmargin{5mm} \onelinecaptionstrue \captionstyle{normal}
 \includegraphics[scale=0.40]{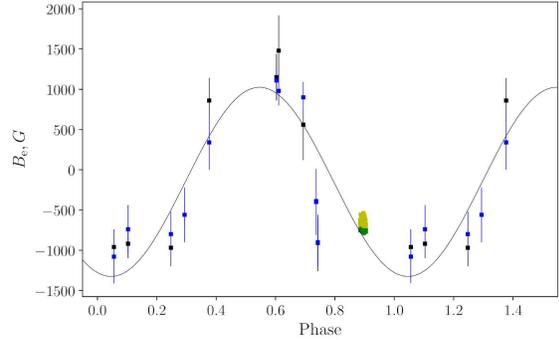}   
 \caption{\small Phase curve of the effective magnetic field $B_e$ of $\alpha^2\,$CVn  with phases corresponding to the rotation period $P=5.43730\,$d.
                 The values of $B_e$ obtained in this work are shown by green (cog method) and yellow (regression method) dots.
          }
 \label{Fig.alf2CVnPhaseCurve}
\end{figure}

Columns 3Ц5 in Table~\ref{Table.MF_Vrad} show the rotation phases of $\alpha^2\,$CVn for rotation periods $P=5.46939\,$d, $P=5.49800\,$d, and $P=5.43730\,$d. The results of our analysis 
show that the period of $P=5.43730\,$d derived from an analysis of TESS satellite observations allows us to describe better the dependence of the effective magnetic field $B_\mathrm{e}$ 
on the rotation phase of the star (phase curve). 

The phase curve for $P=5.43730\,$d is shown in Fig.~\ref{Fig.alf2CVnPhaseCurve}. The values of $B_\mathrm{e}$ taken from Gerth et al.~\cite{Gerth-1999} and Romanyuk 
et al.~\cite{Romanyuk-2016,Romanyuk-2018}, are supplemented by values derived in the present work. The determined longitudinal magnetic field values show a good fit with
the phase curve. The average of all the measured longitudinal magnetic field values is $B_\mathrm{e}=600\pm 56\,$G.

\section{CONCLUSIONS}
\label{s.concl}

In this work we studied the line profile variations in the spectrum of the $\alpha^2\,$CVn star with a 2Ц3 minute  temporal resolution based on spectropolarimetric
BTA observations with the MSS spectrograph. Regular profile variation components were discovered with periods of 5-140~minutes. The presence of longer regular components is also possible.

We determined the rotation period of $\alpha^2\,$CVn using the TESS satellite photometric data, $P=5.43730\pm 0.470\,$d.  
The measured magnetic field corresponds to the magnetic field phase curve for $\alpha^2\,$CVn.

\begin{acknowledgments}
The authors are grateful to I.I. Romanyuk for advices and recommendations which helped to improve the text of the paper.
\end{acknowledgments}

\section*{FUNDING}
A.K. and A.M. acknowledge the support of the RFBR grant 19-02-00311~A. I.Ya. is grateful for the support of this work by the RFBR project no. 19-32-60007. Observations
with the SAO RAS telescopes are carried out with the support of the Ministry of Science and Higher Education of the Russian Federation (including agreement no. 05.619.21.0016,
unique project identifier RFMEFI61919X0016).

\section*{CONFLICT OF INTEREST}
The authors declare no conflict of interest.

 \bibliographystyle{AstroBull}
\bibliography{Kholtygin-2020_alf2CVnE}
\end{document}